\documentclass{article}
\usepackage[utf8]{inputenc}
\usepackage{authblk}
\usepackage{setspace}
\usepackage[margin=1.25in]{geometry}
\usepackage{graphicx}
\graphicspath{ {./figures/} }
\usepackage{subcaption}
\usepackage{amsmath}
\usepackage[labelfont=bf]{caption}
\usepackage[normalem]{ulem}
\usepackage{xcolor}

\newcommand*\mycaption[2]{\caption[#1]{#1#2}}

\usepackage[style=ieee, 
citestyle=numeric-comp,
sorting=none]{biblatex}
\let\oldcaption\caption
\renewcommand\caption[2][]{\oldcaption[#1]{\textbf{#1.} #2}}

\title{Optical Measurement of Photorecombination Time Delays}

\author[1,2,*,$\dag$]{Chunmei Zhang}
\author[1,3,*,$\dag$]{Graham Brown}
\author[1]{Dong Hyuk Ko}
\author[1]{P. B. Corkum}

\affil[1]{Joint Attosecond Science Laboratory, University of Ottawa and National Research Council of Canada, 25 Templeton St, Ottawa, ON K1N 6N5, Canada}
\affil[2]{Beijing Institute of Technology, No. 5, South Street, Zhongguancun, Haidian District, Beijing, China }
\affil[3]{Max Born Institute, Max Born Str. 2a, 12489 Berlin, Germany}
\affil[*]{Corresponding author: chunmei.zhang@bit.edu.cn, graham.brown@uottawa.ca}
\affil[$\dag$]{These authors contributed equally to this work.}

\date{}

\begin{document}

\allowdisplaybreaks

\maketitle

\begin{abstract}
Recollision physics and attosecond pulse generation meld the precision of optics with collision physics. As a follow-up to our previous work, we reveal a new direction for the study of electronic structure and multielectron dynamics by exploiting the collision-physics nature of recollision. We show experimentally that, by perturbing recollision trajectories with an infrared field, photorecombination time delays can be measured entirely optically using the Cooper minimum in argon as an example. In doing so, we demonstrate the relationship between recollision trajectories and the transition moment coupling the ground and continuum states. In particular, we show that recollision trajectories are influenced by their parent ion, while it is commonly assumed they are not. Our work paves the way for the entirely optical measurement of ultrafast electron dynamics and photorecombination delays due to electronic structure, multielectron interaction, and strong-field driven dynamics in complex molecular systems and correlated solid-state systems. 
\end{abstract}


\section*{Introduction}

Recollision consists of three steps \cite{PhysRevLett.71.1994}: (1) in the presence of a strong field, an electron tunnels into the continuum (ionization), (2) the electron is accelerated by the strong field (propagation), and (3) the electron recombines with its parent ion and emits an attosecond pulse (recombination). The spectral phase of attosecond pulses generated through recollision is predominantly shaped by the field-driven continuum dynamics during the propagation step \cite{varjuFCHAP,lewensteinSFA}, but can be significantly shaped by the phase of the transition moment coupling the ground and continuum states during tunnel ionization or recombination \cite{schounAPSACM}. Thus, the measurement of the spectral phase of attosecond pulses can reveal information about the underlying structure of the generation medium through the characterization of the transition moment phase.

The measurement of the spectral phase of attosecond pulses is most commonly accomplished by photoionizing a target atom with an attosecond pulse in the presence of a weak infrared field \cite{paulOTAP,itataniASC,cattaneoCRASC}. In this regime, known as attosecond \emph{ex situ} measurement, the phase of the attosecond pulse and, thus, the transition moment of the generation atom is imprinted onto the photoelectron spectrum from the target atom and this can be measured by monitoring the variation of the photoelectron spectrum with the delay between the attosecond pulse and infrared field.
	
An experimentally simpler alternative form of attosecond measurement was first proposed in 2006 \cite{inSitu3} and involves the perturbation of the recollision process which generates an attosecond pulse with a weak infrared field. As the delay between the weak infrared field and the field driving recollision is varied, the attosecond pulse intensity spectrum varies and this variation can be used to measure the spectral phase of the recollision electron up to a constant \cite{inSitu1,inSitu2,inSitu3,inSitu4}. This form of measurement, referred to \emph{in situ} or all-optical measurement, has been applied to several gas-phase high harmonic generation (HHG) experiments extended to solid-state HHG \cite{allOpticalBandStructure} and has become an invaluable tool for the characterization of recollision dynamics \cite{Zhang2016,C6FD00130K,Pedatzur2015,PhysRevA.95.051401}. Previous works \cite{PhysRevA.94.023825,PhysRevA.95.051401}, however, have demonstrated that these all-optical techniques are insensitive to transition moment phase shifts arising from two-centre interference \cite{PhysRevA.94.023825} and the recollision-induced excitation of a shape resonance \cite{PhysRevA.95.051401}. Generally, it has been assumed that all-optical attosecond measurement is insensitive to the transition moment phase and only characterizes the dynamics of the recollision electron in the continuum \cite{Kim2014}.

The insensitivity of all-optical measurement to the transition moment phase follows from the assumption that the transition moment phase plays no role in determining recollision trajectories. Quantitative rescattering theory (QRS) \cite{QRS} formalized this assumption and describes attosecond pulse spectra as the product of the a universal continuum electron wave packet, specific to the generation medium only through the ionization dynamics, and the field-free transition moment of the generation atom in the spectral domain. QRS suggests a means for the efficient calculation of HHG from complex systems with sufficient accuracy \cite{farrellPMCM,Itatani2004} and the mathematical formalism of QRS has reinforced the notion that recollision trajectories are independent of the transition moment phase.

In a recent work \cite{myFirst}, we argued that this assumption is incorrect and argued that the transition moment does play a role in the determination of recollision trajectories. The implication of this is that all-optical measurement should be sensitive to the transition moment phase, which appears to conflict with the findings of previous works which demonstrated otherwise \cite{PhysRevA.94.023825,PhysRevA.95.051401}. This work constitutes an experimental follow-up to our previous work \cite{myFirst}, conclusively demonstrates the sensitivity of attosecond all-optical measurement to the transition moment phase, and confirms that the transition moment phase plays a role in the determination of recollision trajectories. To do this, we perform an attosecond all-optical measurement in argon, which exhibits a well-known Cooper minimum \cite{cooperPIOAS} and corresponding $\pi$-phase jump in its photorecombination cross-section. Our experimental results conclusively show that all-optical techniques are sensitive to the transition moment phase shift around the Cooper minimum in argon. We support these experimental findings using both \emph{ab initio} simulations based on time-dependent density functional theory (TD-DFT) and an extension of the strong field approximation (SFA), the semiclassical model of recollision, which incorporates the transition moment phase \cite{myFirst}. The implications of our results are critically important for the interpretation and design of experimental attosecond measurement and has implications regarding the description of attosecond pulse generation on a fundamental level.

\section*{Experimental Design}

We begin by describing our experimental all-optical measurement and present the measured attosecond pulse spectral intensity and group delay alongside a simulation of the experimental conditions calculated using TD-DFT \cite{maitraTDDFT}. Our experimental measurement of the group delay variation around the Cooper minimum agrees with the simulation of experimental conditions. We then validate our interpretation of the experimental results by simulating perturbed attosecond pulse emission in argon using our theoretical model. We first show the difference in the spectral phase between perturbed and unperturbed attosecond pulse emission, demonstrating that the perturbation-induced phase shift is sensitive to the transition moment phase. We then present the results of a simulated optical measurement which agrees with our experimental results. Finally, we present a physical interpretation of how the Cooper minimum affects recollision trajectories based on the semi-classical description of recollision known as the SFA \cite{lewensteinSFA}. 

For our experiment, we generate an isolated attosecond pulse in argon using polarization gating \cite{oronPG}, which involves superposing left- and right-circularly polarized pulses separated by a small time-delay. This results in a linear polarization within the temporal window where the two polarizations overlap. Since recollision is extremely sensitive to the ellipticity of the driving field \cite{flettnerEDHHG}, attosecond pulse emission is confined to this overlap region resulting in the emission of an isolated attosecond pulse. 

The experimental diagram of our all-optical measurement is shown in Fig. \ref{fig:1} (a). For more experimental details, please see the supplementary material. We focus a 12 fs, polarization-gated driving pulse with a central wavelength of 1.8 \textmu m into an argon gas jet, as shown by the red pulse. We perturb recollision with a second harmonic of the driving pulse delayed by a time $\tau$ that is polarized parallel to the linearly polarized component of the driving field, has a relative intensity less than $2 \times 10^{-4}$ (our procedure for estimating the ratio is discussed in the supplementary material), and is focused into the gas jet at an angle of $\theta_p = 6.9$ mrad with respect to the driving beam. This is depicted by the blue beam in Fig. \ref{fig:1} (a). In the gas jet, the perturbing field modifies the recollision trajectories and imparts a phase shift which depends on the time delay between the driving and perturbing fields and the vertical position of the atom within the gas jet due to the noncollinear geometry. 

The phase shift induced by the perturbing field can be calculated using the SFA as follows: 

\begin{equation}
	\sigma\left(k, t_b, t_r, \phi\right) = \int_{t_b}^{t_r} \left[ k + A(\tau) \right] A_p \left(\tau, \phi\right) d\tau + A_p \left(t_b, \phi\right) \eta \left(k + A\left(t_b\right)\right) - A_p \left(t_b, \phi\right) \eta \left(k + A\left(t_b\right)\right) ,
\end{equation}

\noindent
where $k$ is the continuum electron canonical momentum, $t_b$ is the time of ionization, $t_r$ is the time of recombination, $A(t)$ and $A_p(t, \phi)$ are the driving and perturbing field vector potentials at time $t$ and relative phase $\phi$, $\eta'(k)$ is the gradient of the transition moment phase for continuum state momentum $k$. A full description of the perturbation-induced phase shift is included in the supplementary material, and vector notation has been omitted due to the assumption of a linearly polarized field. For our measurement, the relative phase $\varphi$ between the driving and perturbing field is determined from the time delay between the driving and perturbing fields, the vertical position within the gas jet, and the angle between the driving and perturbing beams \cite{inSitu1}. 

As the time delay between the perturbing and driving fields is varied, the phase shifts of recollision trajectories across the driving field beam front change. This results in a varying far-field angular deflection of each spectral component of the extreme ultra-violet (XUV) emission, which we label as $\theta_{\Omega}(\tau)$. The difference in the phase of this modulation between two energies is proportional to the difference in XUV emission time \cite{inSitu1}. This is depicted in Fig. \ref{fig:1} (c), which shows the resultant spectrograms for photon energies of 60 (top) and 80 eV (bottom). The difference in the modulation phase is proportional to the difference in emission times $\Delta\tau_{\Omega}$ for these energies. We have similar spectrograms for each frequency component. We record 200 attosecond pulse spectra for each delay step to reduce the influence of noise and fluctuations on the measurement. We then group the scans into four individual measurements which we analyze separately to obtain the measured uncertainty.

\section*{Results and Discussion}

\begin{figure*}
	\includegraphics[width=\textwidth]{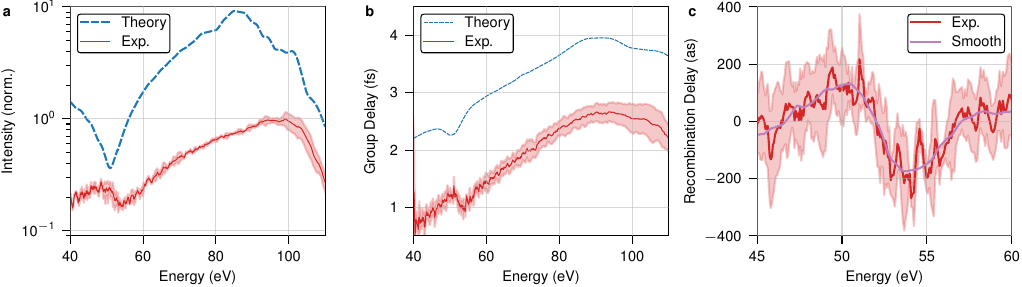}
	\mycaption{\textbf{Experimental All-Optical Measurement in Argon} }{(a) Measured intensity spectrum and (b) group delay for the attosecond pulse generated in argon (solid line) by a 12 fs 1.8 \textmu m driving field with a peak intensity of $1 \times 10^{14}$ W/cm$^{2}$. The dashed lines show the spectrum and group delay from a TD-DFT simulation of the experimental conditions. (c) Difference between the measured group delay and the expected semi-classical result in the spectral region of the Cooper minimum before (red) and after (magenta) smoothing the data. The shaded red region denotes the experimental uncertainty.}
		\label{fig:2}
\end{figure*}

The emission time analysis is performed for each spectral component of the measured attosecond pulse, providing a measurement of the difference in emission time for each spectral component. The group delay of the recollision electron is obtained from the relative delay of each spectral component with respect to that of a fixed energy. Our experimental results and uncertainty are depicted in Fig. \ref{fig:2} by the solid red lines and shaded red regions in all subfigures, respectively. The unperturbed attosecond pulse spectrum is shown in Fig. \ref{fig:2} (a) in red, where the Cooper minimum is observed near 54 eV. Based on the cutoff XUV photon energy of 110 eV, the pulse reaches an peak intensity of $1 \times 10^{14}$ W/cm$^{2}$ within the jet. The attosecond pulse spectrum calculated using a TD-DFT simulation of the unperturbed experimental conditions is shown by the dashed line (scaled $\times 10$ for clarity). Absorbing boundaries \cite{manolopoulosABC} are used beginning at a radius of $F_0 / \omega_0^2$ to suppress long trajectory emission, where $F_0$ is the peak electric field amplitude and $\omega_0$ is the driving field frequency. The position of the Cooper minimum in the TD-DFT simulation, near 51 eV, is lower than the experimental result, near 54 eV. The positions of the Cooper minimum from our simulation and experiment, however, both agree with previously published theoretical \cite{PhysRevA.87.053406,pabstMCCM} and experimental results \cite{PhysRevA.83.053401}. Given the message of this work concerns the qualitative capabilities of all-optical attosecond measurement, it only is the qualitative dynamics of recollision and the effect of the Cooper minimum on recollision trajectories which are important for our work. We use a thin gas jet to minimize the macroscopic contributions due to phase matching and placed the jet before the beam focus. 

The measured group delay is shown in Fig. \ref{fig:2} (b), where the experimental data is shown as a solid red line and the group delay from our TD-DFT simulation of the experimental conditions is shown by a dashed line with a vertical offset for clarity. Both the experimental and theoretical group delay curves are predominantly linear until the cut-off energy near 90 eV, but deviate from linearity near the Cooper minimum. The difference between the experimentally measured group delay and the linear group delay expected from systems without a transition moment phase shift is the photorecombination time delay. Here, we use the group delay from a time-dependent Schr\"{o}dinger equation (TDSE) simulation of a hydrogenic atom with an ionization potential of 15.8 eV (obtained by scaling the nuclear charge) as our reference, which has a slowly varying transition moment phase at large energies \cite{Popruzhenko_2018}. The difference between the measured group delay and the group delay from the simulation of the reference hydrogenic atom calculated with the experimental conditions represents the photorecombination time delay from our experimental measurement and is shown in Fig. \ref{fig:2} (c) in red and a smoothed version of the curve is shown in magenta. The maximum delay difference around the Cooper minimum near 54 eV is 180 as, in agreement with other studies of photorecombination time delays in argon \cite{schounAPSACM}. The positive delay measured at energies below the Cooper minimum are similar to group delay from calculations of photoionization delays in argon \cite{app8030322} and result from the difference in the Coulomb scattering phase of the continuum states populated by the 1s and 3p orbitals of the reference hydrogenic and argon atoms, respectively \cite{Popruzhenko_2018}. 

While our results agree with previous measurements of the photorecombination time delay around the Cooper minimum in argon \cite{schounAPSACM} and our own simulations of the experimental conditions, the energy region around the Cooper minimum lies within an energy range which could exhibit a signal contaminated by second-order diffraction of the high-energy components of our attosecond pulse spectra. We verified our measurement by filtering out the high-energy components of the attosecond pulse spectra and our results remained consistent (details are discussed in the supplementary material). 

To support our experiment, we perform an \emph{ab initio} simulation of an all-optical measurement in argon using TD-DFT \cite{maitraTDDFT}. The details of our implementation of TD-DFT are provided in the supplementary material. We simulate the generation of an isolated attosecond pulse in argon using a single-cycle driving field with a wavelength of 1.8 \textmu m and peak intensity of $1 \times 10^{14}$ W/cm$^{2}$. We use an absorbing boundary \cite{manolopoulosABC} which begins at a radial distance of $0.8 \times F_0 / \omega_0^2$ in order to suppress emission from long trajectories and obtain a clear spectrum. We simulate an all-optical measurement by adding a weak second-harmonic of the driving field with a relative intensity of $10^{-4}$ and scanning the relative phase of the perturbing and driving fields. We do this for argon using TD-DFT and for a reference hydrogenic atom with the same ionization potential (15.8 eV) as the argon atom calculated with DFT and compare these results to isolate the effect of the transition moment phase on the all-optical measurement of attosecond pulse emission from argon.

\begin{figure*}
	\centering
	\includegraphics[width=1.0\textwidth]{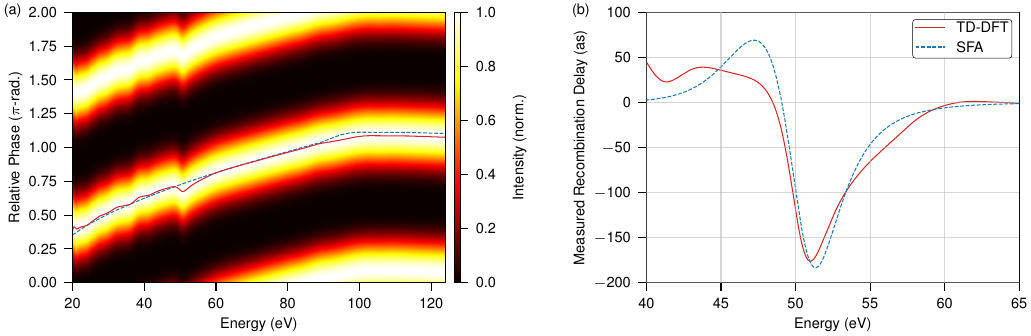}
	\mycaption{\textbf{Simulated all-optical measurement in argon} }{(a) Spectrogram from simulated $\omega-2\omega$ all-optical measurement of attosecond pulse emission in argon calculated using time-dependent density functional theory. The overlaid red and blue lines denote the optimal relative phase between the driving and perturbing fields for each photon energy from all-optical measurements performed in argon and a reference hydrogenic atom with an equivalent ionization potential, respectively. (b) (solid red) The difference between the optimal relative phase for the simulated all-optical measurements in argon calculated using TD-DFT and the reference hydrogenic atom and (dashed blue) the change in the optimal relative phase in a model argon atom due to a Lorentzian $\pi$-phase shift in its recombination moment at 51 eV. The driving field is a single-cycle pulse with wavelength 1.8 \textmu m and peak intensity $1 \times 10^{14}$ W/cm$^{2}$, the perturbing field is a second harmonic of the driving field with relative intensity $10^{-4}$, and absorbing boundaries \cite{manolopoulosABC} are used to suppress long trajectory emission.}
	\label{fig:3}
\end{figure*}

The spectrogram from our simulated all-optical measurement is presented in Fig. \ref{fig:3} (a), which shows the variation of the attosecond pulse intensity spectrum with the relative phase between the driving and perturbing fields normalized for each emitted frequency. The overlaid solid red and dashed blue lines denote the phase which maximizes emission at each frequency from our simulated measurements in argon and the reference hydrogenic atom, respectively. Throughout the entire spectrum, the optimal relative phase from the argon and hydrogenic atom measurements increases with photon energy and plateaus near 100 eV, reflecting the conventional attochirp for short trajectory emission. Below 40 eV, the optimal phase from the argon measurement exhibits several modulations due to multi-channel and multi-pole effects \cite{pabstMCCM}.

At the Cooper minimum near 51 eV, the optimal phase from the argon measurement shows a clear modulation due to the transition moment phase shift. The difference between the optimal phase from the argon and reference hydrogenic atom around the Cooper minimum reveals the how the of the transition moment phase shift around the Cooper minimum affects all-optical measurement. This is shown in Fig. \ref{fig:3} (b) by the solid red curve. The dashed blue curve shows the difference between the measured group delay for an all-optical measurement from a model argon atom with a Lorentzian $\pi$-phase shift in its transition moment and a reference hydrogenic atom calculated using an extension of the SFA which accounts for the transition moment phase shift in the calculation of recollision trajectories \cite{myFirst}. For the SFA calculation, the driving and perturbing fields are sinusoidal fields with the same respective wavelengths and peak intensities as in the TD-DFT calculation and only short trajectory emission is considered. A thorough description of the SFA and all-optical measurement for systems with a transition moment phase shift is provided in the supplementary material.

Despite the comparatively simple model employed by the SFA, the difference in the measured group delay due to the transition moment phase from the TD-DFT simulation and the SFA calculation show qualitative agreement. Both show positive variations in the measured group delay $\sim 50$ as at energies below the Cooper minimum and a maximum decrease in the measured group delay $\sim 180$ as at the Cooper minimum (51 eV). This agreement supports the inclusion of the transition moment phase in the calculation of recollision trajectories in the SFA.

Before discussing the implications of our work, we must first address the apparent contradiction of our results with previous works \cite{PhysRevA.94.023825,PhysRevA.95.051401}. The first of these works \cite{PhysRevA.94.023825} demonstrated theoretically that attosecond all-optical measurement is insensitive to the field-free transition moment phase due to the two-centre interference in a diatomic molecule. In order to obtain a sharp phase shift in the calculated HHG spectra, the first excited state was projected from simulations to reduce the polarizability of the ground state. This projection suppresses the influence of all external fields including the perturbing field used for measurement on the transition moment and forces the HHG spectrum to be shaped predominantly by the field-free transition moment similarly to the formalism suggested by QRS \cite{QRS}. In contrast, our results necessarily include the full response of the system and the transition moment to all external fields. This is expounded upon in the supplementary material.
 
The second of these works demonstrated attosecond all-optical techniques to be insensitive to transition moment phase shifts arising from a  shape resonance in SF$_{6}$ \cite{PhysRevA.95.051401}. The resolution of this disagreement lies in the distinct pathways to recombination in conventional recollision and recollision-induced shape resonance excitation. In the presence of a shape resonance, the continuum electron enters into a localized quasi-bound state prior to recombination to the ground state. The shape resonance in SF$_{6}$ has been shown to be robust against variations of external fields \cite{Manschwetus2015}. Thus, the quasi-bound state is exhibits a much lower polarizability than the electron in the continuum and the influence of the perturbing field on the quasi-bound state is negligible when compared with that on the continuum electron. Consequently, the all-optical measurement of recollision in the presence of a shape resonance is sensitive only to the dynamics occurring prior to the excitation of the shape resonance, which are equivalent to those of conventional recollision. This is also expounded upon in the supplementary material.

It is important to have a qualitative understanding of our results, for which we turn to the semiclassical model of recollision \cite{PhysRevLett.71.1994,lewensteinSFA}. The sensitivity of the all-optical approach to the transition moment phase shift around the Cooper minimum in argon can be explained as a consequence of the spatial structure of the $3p$ orbital wavefunction. Cooper minima arise from a change in the sign of the dipole transition element with continuum state energy, resulting in a phase shift and spectral minimum in the transition moment \cite{cooperPIOAS}. 

Semiclassical recollision trajectories are calculated through the expression for the recollision dipole spectrum calculated using the SFA:

\begin{equation}
	\tilde{D}(\Omega) = - i \int_{- \infty}^{\infty} dt_r \int_{- \infty}^{t_r} dt_b \int dk \hspace{1 mm} d_{0}^{*}(k + A(t_r)) e^{- i S(k, t_b, t_r) - i I_p (t_r - t_b)} F(t_b) d_{0}(k + A(t_b)),
\end{equation}

\noindent 
where $I_p$ is the ionization potential, $A(t)$ and $F(t)$ are the driving field vector potential and electric field at time $t$, $d_0(k)$ is the transition moment between the ground state and continuum state with momentum $k$, and $S(k, t_b, t_r)$ is the semi-classical action. We let $\eta(k)$ denote the phase of the transition moment: 

\begin{equation}
	\eta(k) = \arg\left( d_0(k) \right).
\end{equation}

\noindent 
Eq. (3) is typically solved using saddle-point integration, which involves finding and setting the derivatives of the integrand phase with respect to the integration variables to zero, resulting in a system of equations whose solutions correspond to the dominant contributions of the integrand. The equation obtained from the evaluating derivative of the integrand phase in Eq. (3) with respect to $k$, which is broadly interpreted as a displacement condition for recollision trajectories, including the transition moment phase is given as follows \cite{myFirst}: 

\begin{equation}
	0 = \int_{t_b}^{t_r} \left[ k + A(\tau) \right] d\tau + \eta' \left(k + A(t_b) \right) - \eta' \left(k + A(t_r)\right),
\end{equation}

\noindent 
where $\eta'(k)$ is the gradient of the transition moment phase at momentum $k$. The integration of the kinetic momentum between times $t_b$ and $t_r$ describes the displacement of the recollision electron due to the driving field. For the case of a slowly varying transition moment phase, $| \eta'(k) | \sim 0$ and Eq. (5) states that a recollision electron must return to its position of origin to undergo recollision. When the transition moment phase varies sufficiently rapidly, the second and third terms act as momentum-dependent spatial offsets of the relative positions of ionization and recombination. While the dynamics of recollision and recombination are inherently quantum mechanical and wave-like in nature, the interpretation of Eq. (5) as a displacement condition for semiclassical trajectories and the localization of ionization and recombination during recollision has led to many valuable insights \cite{inSitu1,inSitu2,inSitu3,inSitu4,PhysRevA.99.013412,molSFA1,molSFA2,molSFA3,molSFA4} which we extend here.

\begin{figure*}
	\centering
	\includegraphics[width=0.75\textwidth]{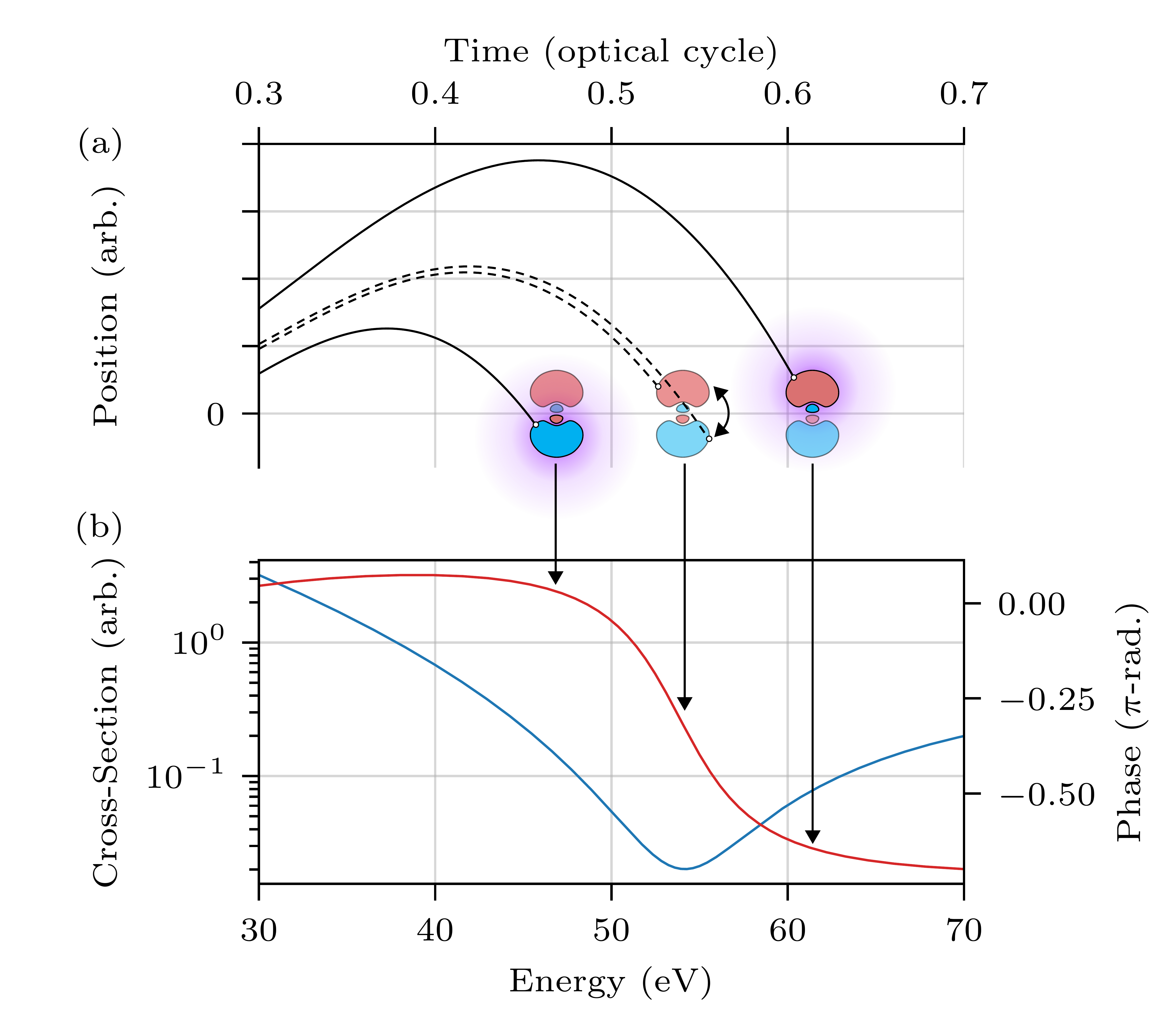}
	\mycaption{\textbf{Qualitative depiction of recollision in argon} }{(a) Recollision trajectories leading to photon emission in argon and (b) the transition moment cross-section (blue, left axis) and phase (red, right axis) in argon. The trajectories in (a) leading to photon emission at a given energy in (b) are denoted by the black vertical arrows. In (a), the $3p_0$ orbital in Ar is overlaid for each trajectory at the time of recombination, denoting the position of recombination with the white circles and the orbital lobe which dominates dipole emission as originally described by Cooper. At energies below (above) the Cooper minimum, the lower (upper) lobe dominates dipole emission. Around the Cooper minimum, the lobe which dominates dipole emission changes, as denoted by the dashed trajectories and the double-headed arrow.}
	\label{fig:4}
\end{figure*}

If we consider the Cooper minimum in argon, the semiclassical interpretation of Eq. (5) is clear: as the recollision electron kinetic energy passes through the Cooper minimum, the displacement condition in Eq. (5) requires that the difference between the positions of ionization and recombination changes according to the gradient of the transition moment phase around the Cooper minimum. This is depicted in Fig. \ref{fig:4}, which shows a qualitative depiction of recollision trajectories as a function of time in (a) and the cross-section (blue, left axis) and transition moment phase (red, right axis) of argon in (b). Due to the linear atto-chirp, the recombination time of a given trajectory can be mapped to an emitted photon energy and this energy is denoted by the vertical black arrows relating Figs. \ref{fig:4} (a) and (b). For a recollision trajectory leading to photon emission below (above) the Cooper minimum, the position of recombination is the lower (upper) lobe of the $3p_0$ orbital. The position of recombination is depicted by the white circles and purple hue, representing XUV emission. At the Cooper minimum, the position of recombination shifts from one lobe to the other as the radial dipole matrix element changes sign \cite{cooperPIOAS}, resulting in the variation of recollision dynamics as observed in our experiment. 

This interpretation is closely related to the localization of ionization and recombination in the molecular SFA \cite{PhysRevA.99.013412}, where the outer lobes of the $3p_0$ orbital play the role of distinct atomic centres in the linear combination of atomic orbital description of molecular systems. As a first-order approximation, the distance travelled by an electron travelling with the kinetic energy leading to photon emission at 54 eV during a time of 180 as is 12 a.u., roughly the spatial extent of the $3p_0$ orbital. A more thorough description of the semi-classical interpretation is provided in a related paper \cite{myFirst}. 

\section*{Conclusion}

To conclude, we have experimentally verified the predictions of our previous work \cite{myFirst} which argued the recollision electron wave packet when observed through attosecond pulse emission is not independent of the transition moment phase. We now stand back even further from the details of the current experiment and consider the close connection between nonlinear optics and photoionization time delay measurements in attosecond physics. Nonlinear optics requires that the nonlinear process, whether perturbative or non-perturbative, returns the system to its initial or a coherently related state. If this were not the case, phase matching would be impossible. In other words, the three-step recollision model requires that photoionization and photorecombination be identical - the recolliding wave packet that produces a pulse is identical to the outgoing wave packet created by that pulse. 

In support of this general principle, we have demonstrated the link between recollision trajectories and the transition moment phase by measuring the photorecombination time delay around the Cooper minimum in argon entirely optically. Compared with traditional forms of attosecond measurement based on photoionization \cite{paulOTAP,itataniASC} that require a sophisticated photoelectron spectrometer in addition to the attosecond pulse generation apparatus \cite{hentschelAM}, optical measurements are much easier to accomplish. Our all-optical approach can be readily generalized to studying multielectron interaction including collision-induced plasma excitation \cite{shinerXe,pabstXe}, Fano resonances \cite{strelkovFR}, and can be extended to gas-phase molecules \cite{biswasPME} and possibly solids \cite{hhgManyBody}. Further, our results suggest a form of electron orbital tomography, wherein the spatial structure of a wavefunction can be inferred from all-optical measurements of recollision. This is expounded upon in a related study which considers the effect of the transition moment phase using the SFA \cite{lewensteinSFA,myFirst}. 

Assessing these differing measurement approaches, measuring time delays by photoionization is encumbered by their sensitivity to many possible sources of delay, including electron dynamics \cite{biswasPME}, molecular structure \cite{vozziTCI}, and dispersive effects during attosecond pulse propagation \cite{cattaneoCRASC}, leading to a lack of clarity regarding the measurement. In contrast, our all-optical measurement is sensitive to the transition moment phase around the Cooper minimum in argon and is predicted to be insensitive to field-free ionic structure \cite{myFirst}.  Other collinear all-optical measurement techniques have been shown to be insensitive to two-centre interference and shape resonances \cite{PhysRevA.94.023825,PhysRevA.95.051401}. Since photoionization and photorecombination experiments measure different things, combining them will allow us to unambiguously isolate the time delays associated with molecular structural and electron dynamic effects. 

Finally, trajectory dynamics and recombination have typically been treated as independent \cite{QRS}. The sensitivity of our all-optical approach, however, demonstrates the limitations of this approximation. The recollision electron wave packet, as observed through the recollision dipole moment, is not independent of its parent ion. Our results have significant implications regarding how recollision is described on a fundamental level. In particular, the independence of continuum electron wave packet when observed through attosecond pulse emission and the generation medium has been used to justify theoretical descriptions of attosecond pulse generation \cite{QRS} and the interpretation of experimental measurements. Our results show that the assumption that recollision trajectories are independent of the transition moment phase is an approximation. The interdependence of recollision trajectories and the source medium must be addressed in the interpretation of all attosecond experiments. This has already been understood in the context of solid-state HHG \cite{jiangEoTDP} and limitations of QRS have been identified in other gas-phase experimental works \cite{Facciala_2018,PhysRevA.99.013412}. Our results clearly demonstrate the limitations of treating recollision trajectories and recombination as independent processes and will pave the way for more efficient measurements of ultrafast electron dynamics in atomic, molecular, and solid-state systems.

\subsection*{Funding}

This research was supported by the United States Air Force Office of Scientific Research (award \#: FA9550-16-1-0109) with contributions from the Canada Foundation for Innovation, the Canada Research Chairs program, Canada's Natural Sciences and Engineering Research Council and the National Research Council of Canada.

\subsection*{Author Contributions}

C.Z. and G.G.B. contributed equally to this work. C.Z. and D.H.K. designed the experiment. C.Z. performed the experiment. G.G.B. performed the numerical and semi-classical theoretical analysis and interpreted the experimental results. G.G.B. and C.Z. analysed the experimental data. G.G.B. and P.B.C. prepared the initial manuscript. All authors contributed in writing the manuscript. C.Z. and P.B.C. supervised the work.

\subsection*{Conflicts of Interest}

 The authors declare that there is no conflict of interest regarding the publication of this article.
 
 \subsection*{Data Availability}
 
 The data that support the findings of this study are available from the corresponding authors upon reasonable request.

\newpage

\section*{Supplementary Material}

\subsection*{Experimental Details}

Figure \ref{fig:s01} shows the detailed experimental setup. The original 1.8 \textmu m beam with 650 \textmu J energy passed a pair of wedges for the control of dispersion and CEP. After a 200 \textmu m BBO crystal, the second order harmonic and fundamental beam are separated with a dichroic mirror. The transmitted 1.8 \textmu m beam was used for generating the polarization gating pulse. The polarization of the reflected second harmonic (SH) beam was controlled to be parallel to the linear portion of the gated pulse by a half waveplate. Eventually, the gating pulse and SH beams are combined and sent to the chamber with a D-shape mirror and focused into the gas jet with the same focusing mirror. The peak intensity of fundamental beam within the jet is estimated with the harmonic cut off. With a cutoff of 110 eV shown in our spectrum in Fig. \ref{fig:2} (a) of the main text, the peak intensity is around $1 \times 10^{14}$ W/cm$^{2}$.

We have calibrated our XUV spectrometer using atomic emission lines as shown in Fig. \ref{fig:s02}. The top figure shows the image on the MCP which records the emission lines induced by a strong 800nm laser beam on a N$_{2}$ target. By integrating the top image vertically, we get the emission line spectrum in the middle figure. The energy of the emission lines can be found from NIST and they are shown with the vertical lines in the bottom figure. Then we compare the emission lines with the spectrum peaks and find pixel indexes of known peaks. The grating equation was used for fitting a curve of the relationship between the image pixel number and photon energy.

The energy of SH beam is measured without resizing by any aperture and it is below the noise level (1 \textmu J) of our energy meter. For making sure the SHG beam covers the whole fundamental beam in the interaction medium, the SHG beam size is apertured from 10 mm to 1-2 mm , lowering the SH energy further ($< 0.15$ \textmu J). With this energy and the focusing geometry, we can estimate the upper bound of the SH  intensity as $0.19 \times 10^{11}$ W/cm$^{2}$. During experiment, the actual intensity ratio is controlled to make sure that it is not too high to get high spatial order HHG and not too low that no spatial modulation is visible.

The spectrum and temporal profile of the driving 1.8 \textmu m laser pulse are shown in Figure \ref{fig:s03} (a) and (b), respectively. Figure \ref{fig:s04} depicts the experimentally acquired attosecond pulse spectra along the centre position of the unperturbed pulse plotted against the relative delay between the perturbing and driving pulse (measured in cycles of the second harmonic). The attosecond pulse spectral intensity at each frequency modulates with the perturbing field delay and the phase of the modulation for each frequency is used to obtain the group delay, as described in the main text. 

To verify the influence of the second-order diffraction of the high-energy components of our attosecond pulse spectra, we repeated the measurement with an Al filter being in the XUV beam path, as shown in Fig. \ref{fig:s01}. Fig. \ref{fig:s05} shows the comparison of the unperturbed attosecond pulse spectra without and with the Al filter and Fig. \ref{fig:s06} shows that of the measured spectrogram. Although the quality of the measurements suffered due to the already low spectral intensity of the spectra around the Cooper minimum, we can see the second order diffraction barely influences the spectrum below the Al filter cutoff. Since second order diffraction would affect our measurement through a slowly varying contribution to the measured group delay over a large energy range, we conclude the phase shift we observe around Cooper minimum is not due to second order diffraction.

\subsection*{Description of Time-Dependent Density Functional Theory Simulations}

For numerical simulations, we use time-dependent density functional theory (TD-DFT) \cite{maitraTDDFT} to describe the argon atom. We approximate the many-electron wavefunction using a set of Kohn-Sham orbitals at position $\mathbf{r}$ and time $t$, $\{\phi_{\alpha} (\mathbf{r},t)\}$, where $\alpha$ denotes the initial orbital angular momentum $l_\alpha$ and magnetic quantum numbers $m_\alpha$ of each state. We express each Kohn-Sham orbital $\phi_\alpha (\mathbf{r},t)$ using a spherical harmonic expansion:

\begin{equation}
	\phi_{\alpha} (\mathbf{r}, t) = \sum_{l = m_{\alpha}}^{N_l} \phi_{\alpha, l, m_{\alpha}}(r, t) Y_{l, m_{\alpha}}(\theta, \phi),
\end{equation}

\noindent
where $r$ is the radial position, $\phi_{\alpha, l, m_{\alpha}} (r,t)$ is the component of the full time-dependent radial Kohn-Sham orbital wavefunction with orbital angular momentum $l$ and magnetic quantum number $m_{\alpha}$, and $\Omega$ is the solid-angle. Since our simulations are restricted to linearly polarized external fields, the magnetic quantum number $m_\alpha$ is conserved throughout simulation.  

The radial grid is discretized using Gauss-Lobatto quadrature \cite{PhysRevA.50.3208} and the field-free Kohn-Sham Hamiltonian is given as follows:

\begin{equation}
	\hat{H}_{ks} = - \frac{1}{2} \frac{\partial^2}{\partial r^2} + \frac{l (l + 1)}{2 r^2} + v_0(r) + \sum_{l_H} v_h^{(l_h)}[n](r) + v_{xc}^{(0)}[n](r) ,
\end{equation}

\noindent
where $v_0 (r)=-Z/r$,  is the ionic potential, $n(\mathbf{r},t)$  is the time-dependent electron density, $v_h^{(l_h)} [n](r,t)$ is the $l_h^{th}$ multipole of the Hartree potential, and $v_{xc}^{(0)} [n](r,t)$ is the zeroth-order multipole of the exchange-correlation potential, which is calculated using the generalized gradient approximation. The potentials $v_h^{(l_h )} [n](r,t)$ and $v_{xc}^{(0)} (r,t)$ are calculated using the electron density. With $g_\alpha$ representing the occupation number of the orbital $\phi_\alpha (\mathbf{r},t)$, the electron density is given as follows:

Explicitly, the multipoles of the Hartree potential in Eq. (S2) are given as follows \cite{jackson_classical_1999}:

\begin{align}
	v_{h}^{(0)}[n](r, t) &= \sum_{\alpha} g_{\alpha} \sum_{l \ge m_{\alpha}} \int_0^{\infty} \frac{\left|\phi_{\alpha, l, m_{\alpha}}(r, t)\right|^2}{r_{>}} dr', \\
	\begin{split}
		v_{h}^{(1)}[n](r, \theta, t) &= \cos(\theta) \sum_{\alpha} g_{\alpha} \sum_{l \ge m_{\alpha}} \int_0^{\infty} \frac{r_{<}}{r_{>}^{2}} \left[ c_{l, m_{\alpha}} \phi_{\alpha, l + 1, m_{\alpha}}^{*}(r, t) + c_{l - 1, m_{\alpha}} \phi_{\alpha, l - 1, m_{\alpha}}^{*}(r, t) \right] \\
		&\hspace{5 mm} \times \phi_{\alpha, l, m_{\alpha}}(r, t) dr' , 
	\end{split}\\
	\begin{split}
		v_{h}^{(2)}[n](r, \theta, t) &= \frac{1}{2} \left(3 \cos^2(\theta) - 1\right) \sum_{\alpha} \sum_{l \ge m_{\alpha}} g_{\alpha} \int_0^{\infty} \frac{r_{<}^2}{r_{>}^3} \left[ q_{l - 2, m_{\alpha}} \phi_{\alpha, l - 2, m_{\alpha}}^{*} (r, t) \right. \\
		&\hspace{5 mm} \left. + p_{l, m_{\alpha}} \phi_{\alpha, l, m_{\alpha}}^{*} (r, t) + q_{l, m_{\alpha}} \phi_{\alpha, l + 2, m_{\alpha}}^{*}(r, t) \right] \phi_{\alpha, l, m_{\alpha}}(r, t) dr' ,
	\end{split}
\end{align}


\noindent 
where $r_{<}, r_{>} = \min(r, r'), \max(r, r')$ and the angular coupling coefficients are given as follows:

\begin{align}
	c_{l,m} &= \sqrt{\frac{(l + 1)^2 - m^2}{(2l + 1)(2 l + 3)}}, \\
	p_{l,m} &= \frac{l (l + 1) - 3 m^2}{(2 l - 1) (2 l + 3)}, \\
	c_{l,m} &= \frac{3}{2 (2 l + 3)} \sqrt{\frac{\left( (l + 1)^2 - m^2 \right) \left( (l + 2)^2 - m^2 \right)}{(2 l + 1)(2 l + 5)}}. 
\end{align}

We use a generalized gradient approximation (GGA), LB94, to describe the exchange-correlation potential which has the following form \cite{PhysRevA.49.2421}:

\begin{equation}
	v_{xc}^{(0)}[n](r, t) = v_{xc}^{(0,LDA)}[n](r, t) - \beta \left(\frac{n(r, t)}{2}\right)^{1/3} \frac{x^2}{1 + 3 \beta \mbox{arcsinh}\left(x\right)},
\end{equation}

\noindent
where $x = | \nabla n(r, t) | / n(r, t)$ and $\beta = 0.05$. Exchange-correlation effects are only kept to the monopole order during time-propagation.

The ground state is calculated self-consistently by numerically diagonalizing the Hamiltonian to obtain the radial Kohn-Sham orbital wavefunctions, calculating the Hartree and exchange-correlation potentials with the new set of Kohn-Sham orbitals, and repeating this procedure until convergence is achieved. 

The time-dependent Kohn-Sham equation describing the evolution of the Kohn-Sham orbital wavefunctions is given as follows:

\begin{equation}
	\begin{split}
		i \frac{\partial \phi_{\alpha, l, m_{\alpha}}}{\partial t} &= \left( r F(t) + v_{h}^{(1)}[n](r, t) \right) \left( c_{l - 1, m_{\alpha}} \phi_{\alpha, l - 1, m_{\alpha}}(r, t) + c_{l, m_{\alpha}} \phi_{\alpha, l + 1, m_{\alpha}}(r, t) \right) \\
		&\hspace{- 5 mm} + v_{h}^{(2)}[n](r, t) \left(q_{l - 1, m_{\alpha}} \phi_{\alpha, l - 2, m_{\alpha}}(r, t) + q_{l, m_{\alpha}} \phi_{\alpha, l + 2, m_{\alpha}}(r, t) \right) \\
		&\hspace{- 20 mm} + \left( - \frac{1}{2} \frac{\partial^2}{\partial r^2} + \frac{l ( l + 1)}{2 r^2} + v_0(r) + v_{h}^{(0)}[n](r, t) + p_{l,m_{\alpha}} v_{h}^{(2)}[n](r, t) + v_{xc}^{(0)}[n](r, t) \right) \phi_{\alpha, l, m_{\alpha}}(r, t)
	\end{split}
\end{equation}

\noindent
Throughout time-propagation, the Hartree potential is expanded to a maximum multipole order of 2 and only the time-dependence of the exchange-correlation zeroth-order multipole is included. The propagator is approximated using the Crank-Nicholson \cite{computationalStrongFieldQuantumDynamics} algorithm and the time-dependent dipole moment $D(t)$ is given by

\begin{equation}
	D(t) = \sum_{\alpha} \sum_{l} \int_0^{\infty} \left[ c_{l, m_{\alpha}} \phi_{\alpha, l + 1, m_{\alpha}}^{*}(r, t) + c_{l - 1, m_{\alpha}} \phi_{\alpha, l - 1, m_{\alpha}}^{*}(r, t) \right] r \phi_{\alpha, l, m_{\alpha}} (r, t) dr. 
\end{equation}

\noindent 
For all time-dependent simulations, a transmission free absorbing boundary \cite{manolopoulosABC} is used. For the time-dependent simulations from a hydrogenic atom displayed in Fig. \ref{fig:2} of the main text, we set the multielectron interaction potentials to zero and scale the nuclear charge for the ionic potential such that the ionization potential matches that of argon (i.e. 15.8 eV). 

\subsection*{Attosecond All-Optical Measurement and the Transition Moment Phase}

In this section, we provide additional details regarding all-optical attosecond measurement and the transition moment phase. We discuss the incorporation of the transition moment phase into the determination of recollision trajectories \cite{myFirst}, the effect of the transition moment phase on the perturbation-induced phase shift in an all-optical measurement, and we address the apparent conflict of our work with previous works \cite{PhysRevA.94.023825,PhysRevA.95.051401}.

\subsubsection*{The Strong Field Approximation Description of All-Optical Measurement with the Transition Moment Phase}

We begin by considering the recollision dipole spectrum calculated using the strong field approximation (SFA). We assume all fields are linearly polarized and omit vector notation such that the dipole spectrum $\tilde{D}(\Omega)$ at frequency $\Omega$ is given as follows:

\begin{equation}
	\begin{split}
		\tilde{D}(\Omega) &= - i  \int_{- \infty}^{\infty} dt_r \int_{- \infty}^{t_r} dt_b \int dk \hspace{1 mm} d_0^*\left(k + A(t_r)\right) e^{- i S(k, t_b, t_r - i I_p (t_r - t_b) } \\
		&\hspace{5 mm} \times F(t_b) d_0\left(k + A(t_b)\right) e^{i \Omega t_r},
	\end{split}
	\label{eq:recollisionDipoleTSFA}
\end{equation}

\noindent
where $k$ is the canonical momentum of the continuum electron, $t_b$ is the time of ionization, $t_r$ is the time of recombination, $- I_p$ is the ground state energy, $A(t)$ and $F(t)$ are the external field vector potential and electric field at time $t$, $d_0(k) = \langle k | \hat{x} | \psi_0 \rangle$ is the dipole transition moment between the ground state $| \psi_0 \rangle$ and continuum state $| k \rangle$ with momentum $k$, and $S(k, t_b, t_r)$ is the semiclassical action \cite{lewensteinSFA}. We let $\Phi_0(k, t_b, t_r)$ denote the component of the integrand phase related to the field-driven continuum dynamics, the evolution of the ground state, and photon emission at frequency $\Omega$:

\begin{equation}
	\Phi_0 (k, t_b, t_r) = S(k, t_b, t_r) + I_p (t_r - t_b) - \Omega t_r,
	\label{eq:zerothOrderPhase}
\end{equation}

\noindent
In addition, we incorporate the phase of the transition moment between the ground state and continuum state with momentum $k$, $\eta(k) = \arg(d_0(k))$. Like in quantitative rescattering theory (QRS) \cite{QRS}, we use the known cross-section and phase to describe the transition moment. With this, the total integrand phase $\Phi(k, t_b, t_r)$ in Eq. (\ref{eq:recollisionDipoleTSFA}) is given as follows:

\begin{equation}
\Phi \left(k, t_b, t_r\right) = \Phi_0\left(k, t_b, t_r\right) + \eta\left( k + A \left(t_b \right) \right) - \eta\left(k + A\left(t_r\right)\right) . 
	\label{eq:totalPhase}
\end{equation}

The integrals in Eq. (\ref{eq:recollisionDipoleTSFA}) are typically solved using saddle-point integration, wherein the integral is approximated as an integral of a Gaussian function situated at the stationary points (referred to as saddle-point solutions) of the integrand phase in Eq. (\ref{eq:totalPhase}). Typically, the transition moment phase is omitted from this saddle-point analysis. As we first proposed in \cite{myFirst}, we now incorporate the transition moment phase into the calculation of the saddle-point solutions. We set the derivatives of Eq. (\ref{eq:totalPhase}) with respect to $k$, $t_b$, and $t_r$ to zero and obtain the following set of saddle-point equations:

\begin{align}
	0 &= \int_{t_b}^{t_r} \left[ k + A(\tau) \right] d\tau + \eta'\left(k + A(t_b)\right) - \eta' \left(k + A(t_r)\right), \label{eq:SPK} \\
	0 &= \frac{\left[ k + A(t_b) \right]^2}{2} + I_p + F(t_b) \eta'\left( k + A(t_b) \right), \label{eq:SPTB} \\
	\Omega &= \frac{\left[ k + A(t_r) \right]^2}{2} + I_p + F(t_r) \eta'\left( k + A(t_r) \right).  \label{eq:SPTR}
\end{align}

Above, the transition moment phase behaves similarly to a spatial offset of the recollision trajectory described by the saddle-point solutions to Eqs. (\ref{eq:SPK}-\ref{eq:SPTR}). In Eq. (\ref{eq:SPK}), the displacement condition for the trajectory requires that the relative positions of ionization and recombination is shifted according to the gradient of the transition moment phase. In Eqs. (\ref{eq:SPTB}) and (\ref{eq:SPTR}), the gradient of the transition moment phase modifies the energy conservation conditions for ionization and recombination, respectively, similarly to a dipole shift in energy due to a spatial displacement equivalent to the gradient of the transition moment phase.

In order to isolate the effect of the transition moment phase on the saddle-point solutions, we treat the transition moment phase as a perturbation in Eqs. (\ref{eq:SPK}-\ref{eq:SPTR}) about the case where $\eta'(k) \equiv 0$ for all $k$, corresponding to the conventional SFA. We let $\lambda$ denote the perturbation-order parameter, $\eta(k) \to \lambda \eta(k)$ in Eqs. (\ref{eq:recollisionDipoleTSFA}) and (\ref{eq:SPK}-\ref{eq:SPTR}), and let $\overline{k}$, $\overline{t}_b$, and $\overline{t}_r$ denote the saddle-point solutions for the case where $\eta'(k) \equiv 0$. We refer to these as the zeroth-order saddle-point solutions. We then expand the saddle-point solutions to second-order in $\lambda$:

\begin{equation}
	\mu = \overline{\mu} + \lambda \Delta{\mu}^{(1)} + \lambda^2 \Delta{\mu}^{(2)}, \label{eq:correctedMu}
\end{equation}

\noindent
with $\mu = k, t_b, t_r$. We substitute the saddle-point solutions described by Eq. (\ref{eq:correctedMu}) into the saddle-point equations in Eqs. (\ref{eq:SPK}-\ref{eq:SPTR}), and expand the result to first-order in $\lambda$. This yields the following set of equations for the first-order corrections to the zeroth-order saddle-point solutions due to the transition moment phase:

\begin{align}
	0 &= \left(\overline{k} + A(\overline{t}_r)\right) \Delta t_r^{(1)} - \left(\overline{k} + A(\overline{t}_b)\right) \Delta t_b^{(1)} + \left( \overline{t}_r - \overline{t}_b\right) \Delta k^{(1)} + \eta'(\overline{k} + A(\overline{t}_b)) - \eta'(\overline{k} + A(\overline{t}_r)), \label{eq:FOCK} \\
	0 &= - \left( \overline{k} +A(\overline{t}_b)\right) F(\overline{t}_b) \Delta t_b^{(1)} + (\overline{k} + A(\overline{t}_b)) \Delta k^{(1)}+F(\overline{t}_b) \eta'(\overline{k} + A(\overline{t}_b)), \\
	0 &= - (\overline{k} + A(\overline{t}_r)) F(\overline{t}_r) \Delta t_r^{(1)} + (\overline{k} + A(\overline{t}_r)) \Delta k^{(1)} + F(\overline{t}_r) \eta'(\overline{k} + A(\overline{t}_r)). 
\end{align}

\noindent
These equations can be solved analytically for $\Delta k^{(1)}, \Delta t_b^{(1)}, \Delta t_r^{(1)}$ \cite{myFirst}:

\begin{align}
	\Delta k^{(1)} &= 0, \label{eq:SPCK0} \\
	\Delta t_b^{(1)} &= \frac{\eta'\left(\overline{k} + A(\overline{t}_b)\right)}{\overline{k} + A(\overline{t}_b)}, \label{eq:SPCTB0} \\
	\Delta t_r^{(1)} &= \frac{\eta'\left(\overline{k} + A(\overline{t}_r)\right)}{\overline{k} + A(\overline{t}_r)} . \label{eq:SPCTR0}
\end{align}

\noindent
For the sake of brevity, the system of equations used to obtain the second-order saddle-point corrections are given in Eqs. (\ref{eq:SPC2EQK}-\ref{eq:SPC2EQTR}) in a subsequent section below.

To demonstrate the effect of the transition moment phase on the recombination time, Fig. \ref{fig:s07} (a) depicts the recombination times calculated for a model argon atom with an ionization potential of 15.8 eV and a Lorentzian $\pi$-phase shift in its transition moment at 54 eV corrected to (solid blue) zeroth-order and (dashed red) second-order with respect to the transition moment phase. The driving field is a sinusoidal field with peak intensity $1 \times 10^{14}$ W/cm$^{2}$ and wavelength $1.8$ \textmu m. Fig. \ref{fig:s07} (b) depicts (solid blue) the first-order change in recombination time for both short and long trajectories and the second-order recombination time correction for (solid red) short and (dashed green) long trajectories due to the transition moment phase.

In an all-optical attosecond measurement, a weak perturbing field with vector potential $A_p (t, \varphi)$ and electric field $F_p (t, \varphi)$ at time $t$ with a relative amplitude $\xi$ and phase $\varphi$ with respect to the driving field is added to the recollision process. The perturbing imparts a phase shift onto the recollision trajectories and, as the relative phase between the driving and perturbing field is varied, the generated attosecond pulse intensity spectrum modulates according to this phase shift. The modulations of the measured attosecond pulse intensity spectra with $\varphi$ can be used to reconstruct the underlying recollision dynamics \cite{inSitu1,inSitu2,inSitu3,inSitu4}.

We calculate the perturbation-induced phase shift, $\sigma(k, t_b, t_r, \varphi)$, by expanding the recollision dipole phase in Eq. (\ref{eq:totalPhase}) to first-order in the relative amplitude of the perturbing field \cite{inSitu3}:

\begin{equation}
	\begin{split}
		\sigma (k, t_b, t_r, \varphi) &= \int_{t_b}^{t_r} \left[ k + A(\tau) \right] A_p (\tau, \varphi) d\tau + A_p (t_b, \varphi) \eta' (k + A(t_b)) - A_p (t_r, \varphi) \eta' (k + A(t_r)) .
	\end{split}
	\label{eq:sigmaFull}
\end{equation}

\noindent
We now expand Eq. (\ref{eq:sigmaFull}) through second-order in the transition moment phase.  Since the transition moment phase is most important during recombination, we omit the contribution from the transition moment phase at the time of birth and the corrections to the perturbation-induced phase shift due to the corrections to the time of ionization and canonical momentum for brevity. This does not affect our conclusions. The expression we obtain for $\sigma(k, t_b, t_r, \varphi)$ corrected to second-order in the transition moment phase is the following:

\begin{equation}
	\begin{split}
		\sigma(k, t_b, t_r, \varphi) &= \sigma^{(0)} (\overline{k}, \overline{t}_b, \overline{t}_r, \varphi) + \Delta t_r^{(2)} \left( \overline{k} + A(\overline{t}_r) \right) A_p (\overline{t}_r, \varphi) \\
		&\hspace{ 5 mm}  + \Delta t_r^{(1)} \left( F_p (\overline{t}_r, \varphi) \eta' (\overline{k} + A(\overline{t}_r)) - A_p (\overline{t}_r, \varphi) F(\overline{t}_r) \eta''(\overline{k} + A(\overline{t}_r)) \right) + \mathcal{O}\left(\lambda^3\right).
	\end{split}
	\label{eq:pertShiftChange}
\end{equation}

\noindent
where $\sigma^{(0)}(\overline{k}, \overline{t}_b, \overline{t}_r, \varphi)$ is the perturbation-induced phase shift from an equivalent atom without a rapidly transition moment phase.

The spectrogram \cite{inSitu1,inSitu3} from an all-optical measurement in a model argon atom with an ionization potential of 15.8 eV and a Lorentzian $\pi$-phase shift in its transition moment for photon emission at 54 eV calculated using the SFA model presented above is shown in Fig. \ref{fig:s08} (a). For this calculation, we used a sinusoidal driving field with wavelength 1.8 \textmu m and peak intensity of $1 \times 10^{14}$ W/cm$^{2}$ and a second-harmonic perturbing field with a relative intensity of $10^{-4}$. The overlaid dashed green and solid red lines denote the optimal relative phase between the driving and perturbing fields for emission at each photon energy with and without the transition moment phase, respectively. Fig. \ref{fig:s08} (b) shows the difference between the group delay measurements with and without the transition moment phase shift. This result qualitatively agree with the experimental results in Fig. \ref{fig:2} and the results presented in Fig. \ref{fig:3} (b), where they are shown along side the difference in measured group delay from simulated all-optical measurements in argon and a reference hydrogenic atom.

\subsubsection*{Comparison with Previous Works}

As stated in the main text, our experimental and theoretical results appear to disagree with the theoretical results of \cite{PhysRevA.94.023825}, which showed attosecond all-optical measurement to be insensitive to the transition moment phase shift in a one-dimensional model of H$_{2}^{+}$ resulting from two-centre interference. A subsequent experimental study showed all-optical techniques to be insensitive to the phase shift due to a shape resonance in SF$_{6}$. We address both these apparent discrepancies and show our results are consistent with these findings in the subsequent two sections.

\subsubsection*{All-Optical Measurement and Two-Centre Interference} 

In \cite{PhysRevA.94.023825}, the sensitivity of attosecond all-optical and \emph{ex situ} measurement techniques to the transition moment phase shift arising from two-centre interference in a one-dimensional model of H$_{2}^{+}$ were compared. The results indicated that, while attosecond \emph{ex situ} measurement was clearly sensitive to the transition moment phase, all-optical measurement appeared to exhibit no sensitivity to the transition moment. While these conclusions seem clear and decisive, a subtle detail of the simulation design in \cite{PhysRevA.94.023825} is critically important in determining the sensitivity of all-optical measurement to the transition moment phase. To demonstrate how our results are consistent with those in \cite{PhysRevA.94.023825}, we first discuss how the transition moment manifests during attosecond pulse generation in a diatomic molecule.

The field-free transition moment of the diatomic molecular system studied in \cite{PhysRevA.94.023825} exhibits a narrow spectral minimum and discontinuous $\pi$-phase jump. Attosecond pulse spectra generated from  this molecule, however, exhibit a broad and shallow spectral minimum \cite{PhysRevA.99.013412}. The origin of this broadening and the shaping of the observed transition moment phase shift has been identified as being related to the influence of the external field on the ground state, the modified trajectory dynamics due to the internuclear separation, and is related to the influence of the first-excited state during recombination \cite{PhysRevA.87.043424,PhysRevA.89.043407,PhysRevA.99.013412}. Due to this broadening, the phase shift across the spectral minimum varies slowly with emitted photon energy and gives rise to a small variation in the attosecond pulse group delay, which may be difficult to identify in simulations of both attosecond \emph{ex situ} and all-optical measurement.

In \cite{PhysRevA.94.023825}, the first excited state was projected from the HHG simulations used to assess the sensitivity of all-optical measurement to the transition moment phase. This projection reduces the polarizability of the system and forces HHG spectra to be predominantly shaped by the field-free transition moment, thereby imprinting a narrow spectral minimum and phase shift onto emitted HHG spectra. This narrow phase shift leads to a large group delay variation in the emitted HHG spectra which is easily observable using attosecond \emph{ex situ} techniques because these techniques measure the emitted radiation directly. Attosecond all-optical measurement, however, is performed during the attosecond pulse generation process itself. The results of an all-optical measurement and the dynamics of the electron wavefunction are inherently interdependent and everything which influences the electron wavefunction must be accounted for in the interpretation of an all-optical measurement. In this regard, the effect of the projection of the first excited state on an all-optical measurement must be accounted for.

To demonstrate the importance of the projection of the first excited state in an all-optical measurement, we let $\psi_0(x)$ and $\mathcal{D}(k)$ denote the ground state wavefunction at position $x$ and its recombination moment including its complex phase at continuum state momentum $k$ for a given system. The effect of the perturbing field on the recombination moment can be found by expanding the recombination moment to first-order in the relative amplitude of the perturbing field, $\xi$:

\begin{equation}
	\mathcal{D}\left(k + A(t_r) + A_p (t_r, \varphi)\right) = \mathcal{D}\left(k + A(t_r)\right) +  A_p \left(t_r, \varphi \right) \mathcal{D}'\left(k + A(t_r)\right) + \mathcal{O}\left(\xi^2\right).
\end{equation}

\noindent
Within the plane-wave approximation used by the conventional SFA, the momentum-derivative of the recombination moment can be expressed as follows:

\begin{equation}
	\mathcal{D}'\left(k\right) = \int_{- \infty}^{\infty} x \tilde{\psi}_1(x) e^{i k x},
	\label{eq:generalTMR}
\end{equation}

\noindent
where $\tilde{\psi}_1(x) = x \psi_0(x)$. For the case of the considered diatomic molecular system studied in \cite{PhysRevA.94.023825}, the projection of the normalized state $\tilde{\psi}_1(x)$ onto the true first excited $\psi_1(x)$ state is near unity ($\langle \tilde{\psi}_1 | \psi_1 \rangle = 0.99$). That is, within the plane-wave approximation for the continuum states, the projection of the first excited nearly exactly cancels the effect of the perturbing field.

While the plane-wave approximation provides a simple picture, it is known that it does not accurately describe two-centre interference \cite{PhysRevA.99.013412}. A similar conclusion can be drawn, however, by considering the true continuum eigenstates of the system. Fig. \ref{fig:s09} depicts the (solid blue) field-free recombination cross-section of the ground state, (dashed red) its derivative with respect to the continuum state momentum, and (dot-dash green) the recombination cross-section of the first-excited state as a function of continuum state momentum. Around the spectral minimum, the momentum-derivative of the ground state recombination moment and the transition moment of the first excited state are nearly equal, demonstrating the importance of the first-excited state on the perturbation of the transition moment around the spectral minimum by the perturbing field.

The projection of the first excited state suppresses the influence of the perturbing field on the transition moment in Eq. (\ref{eq:generalTMR}) and artificially restricts its influence predominantly to the perturbation-induced phase shift described by Eq. (\ref{eq:pertShiftChange}). In a previous work \cite{myFirst}, we showed that the perturbation-induced phase shift is insensitive to the field-free ionic structure of a diatomic molecule through second-order in the internuclear separation when described using the molecular SFA \cite{PhysRevA.99.013412}. Thus, we may say that all-optical techniques are insensitive to the field-free ionic structure when the first excited state is projected out from simulations. If the first excited state is retained in simulations, however, the transition moment should be observable in an all-optical measurement according to Eq. (\ref{eq:generalTMR}).

To check this, we simulate all-optical measurements in the same molecule studied in \cite{PhysRevA.94.023825} where we either retain and project out the first excited state. While retaining the first excited state in numerical simulation will result in a phase shift which varies slowly with the emitted photon energy and may be difficult to identify, retaining the first excited state preserves the full effect of the perturbing field on the transition moment and constitutes a complete test of the sensitivity of all-optical techniques to the transition moment. In order to identify such a slowly varying phase shift in our simulated measurements, we also simulate an all-optical measurement from a one-dimensional reference atom with an equivalent ionization potential and compare the measurement results from both systems.

For these measurements, we simulate the generation of an isolated attosecond pulse using a single-cycle driving field with wavelength 1.8 \textmu m and peak intensity of $1 \times 10^{14}$ W/cm$^{2}$ and use an absorbing boundary \cite{manolopoulosABC} to suppress long trajectory emission and obtain clear attosecond intensity spectra. We simulate the all-optical measurements by introducing a second-harmonic perturbing field with a relative intensity of $10^{-4}$ and scan the relative phase between the driving and perturbing fields. The phase of the modulation of each frequency component's spectral intensity with the relative phase between the driving and perturbing fields is then used as a measurement of the recollision electron group delay.

The results of the measurements for the case where we retain the first excited state in the simulation are shown in Fig. \ref{fig:s10}. Fig. \ref{fig:s10} (a) depicts the spectrogram obtained from the all-optical measurement of the attosecond pulse emission from the diatomic molecule where we retain the first excited state in the simulation. A similar spectrogram (not shown) is obtained for the simulated measurement from the reference atomic system. At first glance, the spectrogram from the diatomic molecule appears identical to that expected from the reference atomic system. As stated above, however, the spectral minimum is broadened due to the modified recombination dipole moment described in \cite{PhysRevA.99.013412}. Consequently, the phase shift around the spectral minimum varies slowly with photon energy and may be difficult to identify without an appropriate reference.

To determine whether the all-optical measurement is sensitive to the transition moment phase, we calculate the difference in the measured group delay between the measurements from the molecular and reference atomic systems. The results of this are shown in Fig. \ref{fig:s10} (b), which shows (blue, left axis) the attosecond pulse intensity spectrum from the diatomic molecular system and (red, right axis) the difference in the measured group delay between the measurements from the diatomic molecular and reference atomic systems. A broad variation of $\sim 15$ as in the measured group delay difference between the molecular and reference atomic systems is visible around the spectral minimum, similar to the change in recombination time calculated using Eq. (\ref{eq:SPCTR0}) when describing the molecular system as a linear combination of atomic orbitals \cite{PhysRevA.99.013412}.

We now consider the case where we project out the first excited state and simulate an all-optical measurement with identical parameters to those in Fig. \ref{fig:s10}. Since we are considering an isolated attosecond pulse, the spectral resolution of such a measurement will be sufficiently precise so as to reveal the importance of higher-order corrections to the perturbation-induced phase shift in Eq. (\ref{eq:pertShiftChange}) around the spectral minimum. The results of this simulated measurement are shown in Fig. \ref{fig:s11}. Again, the spectrogram in Fig. \ref{fig:s11} (a) resembles that expected from the reference atom (not shown) and the modulation phase as a function of photon energy (solid red line) resembles the conventional attochirp. At the position of the spectral minimum near 46 eV, however, the modulation phase exhibits a small variation. This variation is due to higher-order corrections to the transition moment and perturbation-induced phase shift. Fig. \ref{fig:s11} (b) depicts (solid blue, left axis) the intensity spectrum, (solid red, right axis) calculated group delay, and (solid green) measured group delay from the simulated measurement. While the large variation in the calculated group delay around the spectral minimum is absent in the measurement result, we show in the inset the difference between the measured group delay from the diatomic molecule where we project out the first excited state and the reference atomic system between photon energies of 40 and 50 eV. A small variation of $\sim 25$ as is observed around the spectral minimum due to the aforementioned higher-order corrections to the transition moment and perturbation-induced phase shift.

These results conclusively show that our experimental and simulated measurements of the Cooper minimum phase in argon are consistent with the results of \cite{PhysRevA.94.023825} because we retain the full dynamics of the system. In particular, we retain the full effect of the perturbing field on the transition moment in our simulations and experimental analysis. Our analysis shows that, while all-optical measurement is largely insensitive to the field-free transition moment and ionic structure, the changes in the recollision trajectory dynamics due to the transition moment phase are observable using all-optical techniques. If we consider the ground state $3p_0$ orbital in argon as a two-centre system with each orbital lobe constituting a centre of recombination (where the separation between the peaks of the outer lobes is $\sim 4$ a.u.), our results regarding two-centre can be directly related to the analysis in this section.

\subsubsection*{All-Optical Measurement and the Shape Resonance in SF$_{6}$}

We now consider the results of \cite{PhysRevA.95.051401}, which show attosecond all-optical measurement to be insensitive to the phase shift due to the excitation of a shape resonance in SF$_{6}$. When a shape resonance is excited during recollision prior to recombination, the recollision electron enters into a localized quasi-bound state $| \Theta \rangle$ due to the shape of the ionic potential. If we consider the shape resonance in SF$_{6}$ studied in \cite{PhysRevA.95.051401}, the shape resonance is very robust with respect to the driving field intensity \cite{Manschwetus2015}. This implies the influence of the weak perturbing field on the excited state is negligible.

With this, dipole emission spectrum at frequency $\Omega$ for the recollision processes involving a shape resonance is then given as follows:

\begin{equation}
	\begin{split}
		\tilde{D}(\Omega) &= - i \int dk \int_{- \infty}^{\infty} dt_r \int_{- \infty}^{t_r} dt_{sr} \int_{- \infty}^{t_{sr}} dt_b \hspace{1 mm} d_{0, \Theta}^* d_{\Theta}(k + A(t_{sr})) e^{- i S(k, t_b, t_{sr}) - i E_{\Theta} (t_r - t_{sr})  - i I_p (t_{sr} - t_b)} \\
		&\hspace{5 mm} \times F(t_b) d_0(k + A(t_b)) e^{i \Omega t_r},
	\end{split}
	\label{eq:shapeResonance}
\end{equation}

\noindent 
where $t_b$ is the time of ionization, $t_{sr}$ is the time of excitation of the shape resonance, $t_r$ is the time of recombination to the ground state, $E_{\Theta}$ is the energy of the quasi-bound state, $k$ is the canonical momentum, $d_{0, \Theta}$ is the transition dipole between the ground and quasi-bound state, $d_0(k)$ is the transition dipole between the ground state the continuum state with momentum $k$, and $d_{\Theta}(k) = \langle \Theta | \hat{H} | k \rangle$ is the transition dipole between the continuum state with momentum $k$ and the quasi-bound state $| \Theta \rangle$, and $F(t)$ and $A(t)$ are the external electric field and vector potential at time $t$. The order of integration in Eq. (\ref{eq:shapeResonance}) can changed, yielding the following:

\begin{equation}
	\begin{split}
		\tilde{D}(\Omega) &\propto \frac{1}{\Gamma / 2 - i (\Omega - E_{\xi} - I_p)} \int_{- \infty}^{\infty} dt_{sr} \int_{- \infty}^{t_{sr}} dt_b \int dk \hspace{1 mm} d_{sr}(k + A(t_{sr})) e^{- i S(k, t_b, t_{sr}) - i I_p (t_{sr} - t_b)} \\
		&\hspace{5 mm} \times F(t_b) d_0(k + A(t_b)) e^{i \Omega t_{sr}} \\
		&= \frac{1}{\Gamma / 2 - i (\Omega - E_{\xi} - I_p)} \tilde{D}_0 \left(\Omega\right),
	\end{split}
\end{equation}

\noindent 
where $\Gamma$ is the width of the shape resonance. This expression shows that the calculated spectrum for a recollision process involving the excitation of a shape resonance before recombination to the ground state can be factorized into a term describing a conventional recollision process $\tilde{D}_0(\Omega)$ which finishes in the quasi-bound state and an amplitude and phase modulation involving the transition from the quasi-bound state to the ground state. Since the integrand phase in $\tilde{D}_0(\Omega)$ is equivalent to that the dipole integrand phase describing recollision in a simple atom with $t_r \to t_{sr}$, the recollision dynamics and, thus, the results of an all-optical measurement, are equivalent to that of a simple atom. This description results from the robustness of the shape resonance with respect to the laser as demonstrated in \cite{Manschwetus2015} and is consistent with previous works studying recollision in systems with autoionizing shape resonances \cite{PhysRevA.89.053833}. Thus, the SFA does not predict that all-optical measurement is sensitive to the phase dynamics around the shape resonance in SF$_{6}$.

\subsubsection*{Second-Order Saddle-Point Corrections}

In this section, we show the system of equations used to obtain the second-order saddle-point corrections and their solutions explicitly. After substituting the first-order saddle-point corrections in Eqs. (\ref{eq:SPCK0}-\ref{eq:SPCTR0}) into the saddle-point equations in Eqs. (\ref{eq:SPK}-\ref{eq:SPTR}), we obtain a set of three equations for the second-order saddle-point corrections $\Delta k^{(2)}$, $\Delta t_{b}^{(2)}$, and $\Delta t_{r}^{(2)}$:

\begin{align}
	\begin{split}
		0 &= \left( \overline{k} + A(\overline{t}_r) \right) \Delta t_r^{(2)} - \left(\overline{k} + A(\overline{t}_b) \right) \Delta t_b^{(2)} + \left( \overline{t}_r - \overline{t}_b\right) \Delta k^{(2)} + \frac{F(\overline{t}_b) \eta'^2(\overline{k} + A(\overline{t}_b))}{2 (\overline{k} + A(\overline{t}_b))^2} \\
		&\hspace{5 mm} - \frac{F(\overline{t}_r) \eta'^2 (\overline{k} + A(\overline{t}_r))}{2 (\overline{k} + A(\overline{t}_r))^2}  - \frac{F(\overline{t}_b) \eta'(\overline{k} + A(\overline{t}_b)) \eta''(\overline{k} + A(\overline{t}_b))}{\overline{k} + A(\overline{t}_b)} \\
		&\hspace{10 mm} + \frac{F(\overline{t}_r) \eta'(\overline{k} + A(\overline{t}_r)) \eta''(\overline{k} + A(\overline{t}_r))}{\overline{k} + A(\overline{t}_r)}, \label{eq:SPC2EQK}
	\end{split} \\
	\begin{split}
		0 &= - \left( \overline{k} + A(\overline{t}_b) \right) F(\overline{t}_b) \Delta t_b^{(2)} + \left( \overline{k} + A(\overline{t}_b) \right) \Delta k^{(2)} + \frac{\omega_0^2 A(\overline{t}_b) \eta'^2(\overline{k} + A(\overline{t}_b))}{2 (\overline{k} + A(\overline{t}_b))} \\
		&\hspace{5 mm} + \frac{F^2(\overline{t}_b) \eta'^2(\overline{k} + A(\overline{t}_b))}{2 (\overline{k} + A(\overline{t}_b))^2}  - \frac{F^2(\overline{t}_b) \eta'(\overline{k} + A(\overline{t}_b)) \eta''(\overline{k} + A(\overline{t}_b))}{\overline{k} + A(\overline{t}_b)}, \label{eq:SPC2EQTB}
	\end{split}\\
	\begin{split}
		0 &= - \left( \overline{k} + A(\overline{t}_r) \right) F(\overline{t}_r) \Delta t_r^{(2)} + \left( \overline{k} + A(\overline{t}_r) \right) \Delta k^{(2)} + \frac{\omega_0^2 A(\overline{t}_r) \eta'^2(\overline{k} + A(\overline{t}_r))}{2 (\overline{k} + A(\overline{t}_r))} \\
		&\hspace{5 mm} + \frac{F^2(\overline{t}_r) \eta'^2(\overline{k} + A(\overline{t}_r))}{2 (\overline{k} + A(\overline{t}_r))^2} - \frac{F^2(\overline{t}_r) \eta'(\overline{k} + A(\overline{t}_r)) \eta''(\overline{k} + A(\overline{t}_r))}{\overline{k} + A(\overline{t}_r)}. \label{eq:SPC2EQTR}
	\end{split} 
\end{align}

\clearpage
\break

\begin{figure*}
	\includegraphics[width=\textwidth]{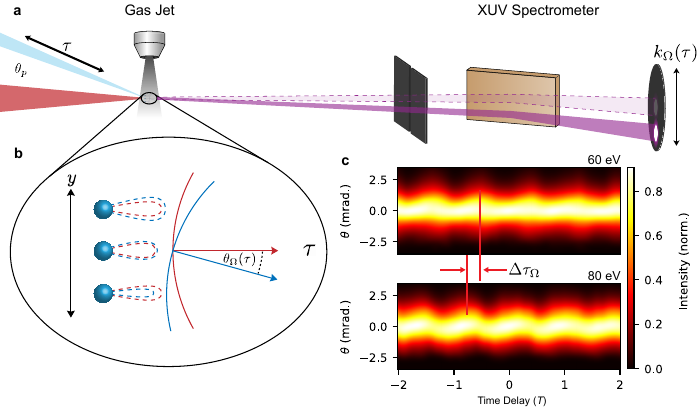}
	\mycaption{\textbf{Experimental Diagram} }{(a) The polarization-gated driving pulse (red) and perturbing pulse delayed by time $\tau$ (blue) are focused into an argon gas jet. The perturbing pulse has a relative intensity less than $10^{-4}$ and incident angle of $\theta_p = 6.9$ mrad with respect to the driving field. (b) The perturbation induces a delay dependent modulation of the electron trajectories and XUV wavefronts (red to blue) across the driving field beam front. This results in a deflection of the XUV beam in the far-field by a delay-dependent angle $\theta_{\Omega}(\tau)$ which is recorded in the XUV spectrometer. (c) The deflection of the XUV emission is recorded over a range of delays (in units of the perturbing field period $T_p$) and spectrally resolved. The resultant spectrograms for XUV energies of 60 (top) and 80 (bottom) eV are shown. The phase difference in the deflection modulation phase is proportional to the difference in group delay, $\Delta \tau_{\Omega}$ between the two energies. }	
	\label{fig:1}
\end{figure*}

\clearpage
\break

\clearpage
\break

\clearpage
\break

\setcounter{figure}{0}
\renewcommand{\thefigure}{S\arabic{figure}}

\clearpage
\break

\begin{figure*}
	\includegraphics[width=\textwidth]{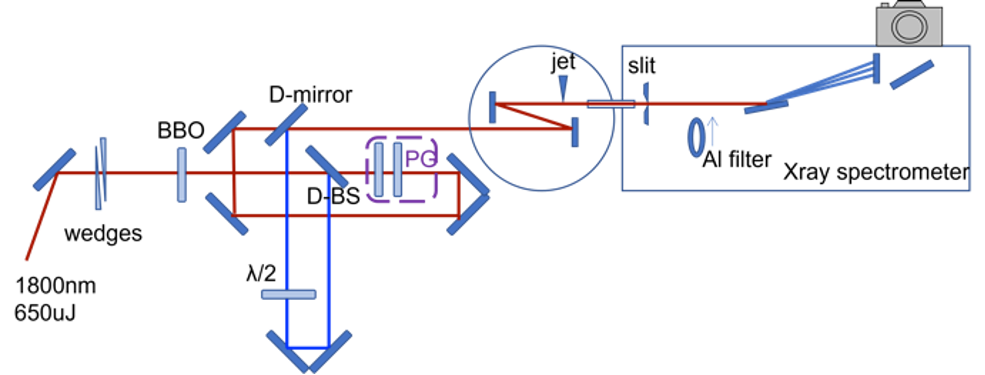}
	\mycaption{\textbf{Optical setup of all optical measurement} }{$\lambda / 2$: half-waveplate; PG: polarization gating setup; D-BS: dichroic mirror; D-mirror: D-shape mirror. The vacuum chamber containing the gas jet (denoted as jet) and the XUV spectrometer is depicted as the area enclosed by the blue circle and rectangle. The differential pumping tube connecting the vacuum chambers containing the gas jet the XUV spectrometer is depicted as the small blue rectangle. The aluminum filter used to obtain the filtered attosecond pulse spectra depicted in Figs. S4 and S5 is denoted as Al filter. }
	\label{fig:s01}
\end{figure*}

\clearpage
\break

\begin{figure*}
	\centering
	\includegraphics[width=0.8\textwidth]{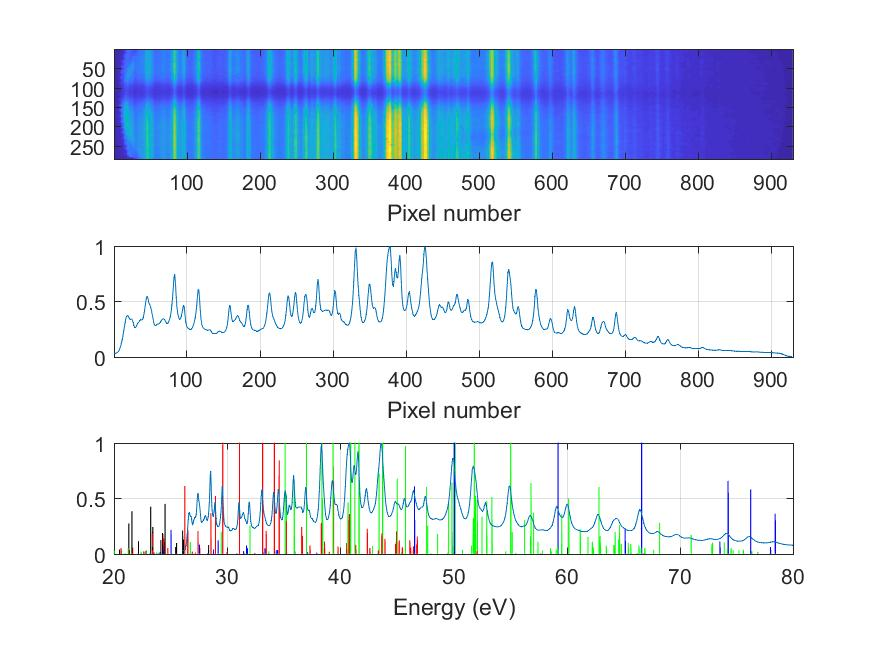}
	\mycaption{\textbf{Calibration using N2 emission lines} }{(top) Emission line image on MCP, (middle) Vertically integrated spectrum, (bottom) Comparison emission lines and spectrum after the calibration.}
	\label{fig:s02}
\end{figure*}

\clearpage
\break

\begin{figure*}
	\includegraphics[width=\textwidth]{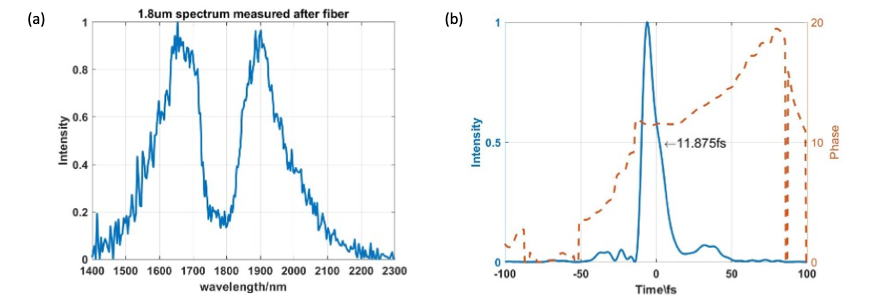}
	\mycaption{\textbf{Driving field spectrum and pulse duration} }{(a) The driving field spectrum at the exit point of the hollow-core fibre. (b) The temporal profile of the driving field pulse after compression of the pulse after propagation through the hollow core fibre.}
	\label{fig:s03}
\end{figure*}

\clearpage
\break

\begin{figure*}
	\centering
	\includegraphics[width=0.75\textwidth]{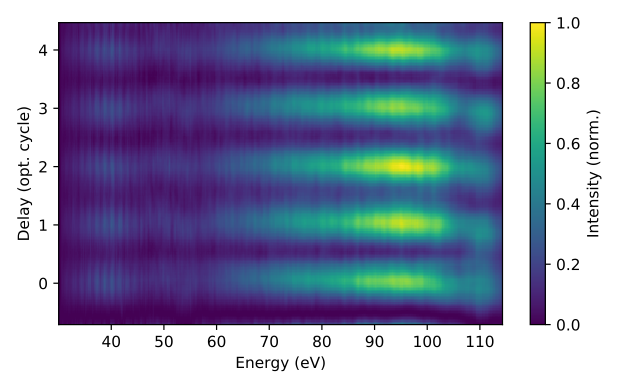}
	\mycaption{\textbf{All-optical measurement on-axis spectrogram} }{The attosecond pulse spectra along the centre position of the unperturbed pulse plotted against the perturbing-driving pulse delay.}
	\label{fig:s04}
\end{figure*}

\clearpage
\break

\begin{figure*}
	\includegraphics[width=\textwidth]{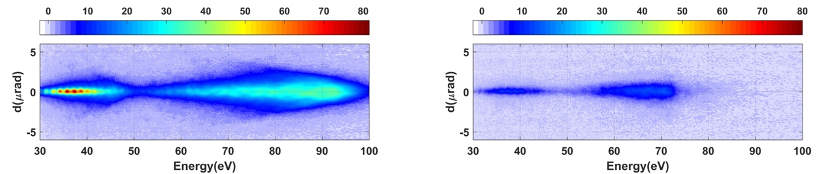}
	\mycaption{\textbf{Attosecond pulse spectra with and without a spectral filter} }{The unperturbed attosecond pulse spectra without and with an Al filter. }
	\label{fig:s05}
\end{figure*}

\clearpage
\break

\begin{figure*}
	\includegraphics[width=\textwidth]{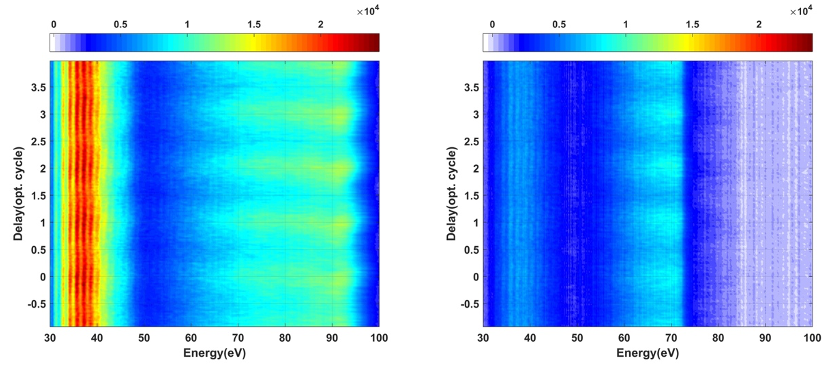}
	\mycaption{\textbf{Spectrograms with and without a spectral filter} }{The attosecond pulse spectra along the centre position of the unperturbed pulse plotted against the perturbing-driving pulse delay without and with an Al filter}
	\label{fig:s06}
\end{figure*}

\clearpage

\begin{figure*}
	\includegraphics[width=\textwidth]{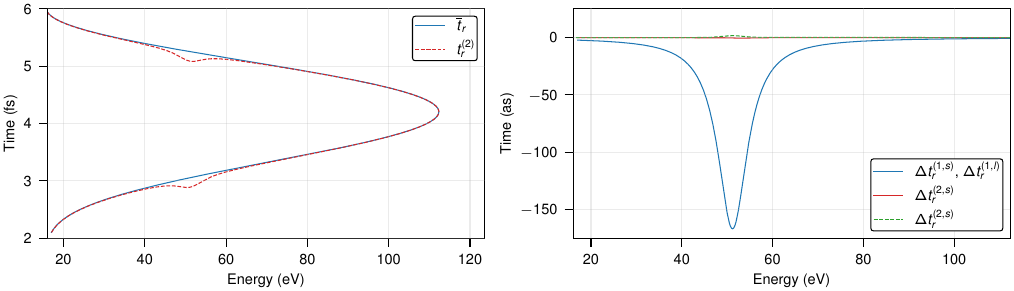}
	\mycaption{\textbf{Recombination Time with the Transition Moment Phase} }{(a) The (solid blue) zeroth-order and (dashed red) second-order recombination times for the short and long trajectories calculated with a Lorentzian phase shift centred at an emitted photon energy of 54 eV. The atom has an ionization potential of 15.8 eV and the driving field is a sinusoidal field with wavelength 1.8 \textmu m and a peak intensity of $1 \times 10^{14}$ W/cm$^{2}$.}
	\label{fig:s07}
\end{figure*}

\clearpage

\begin{figure*}
	\centering
	\includegraphics[width=\textwidth]{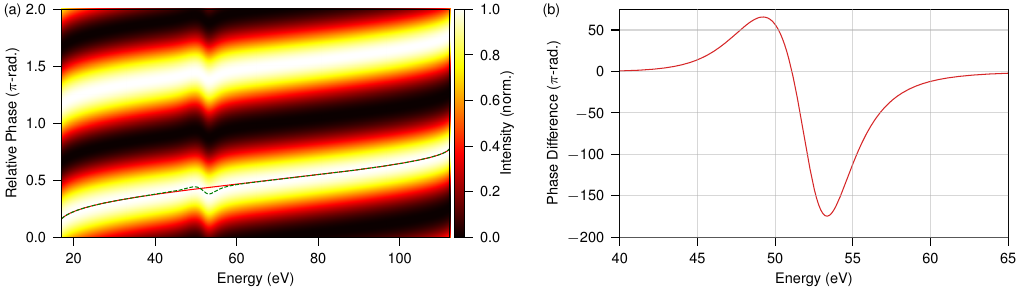}
	\mycaption{\textbf{All-Optical Measurement in argon calculated using the SFA} }{(a) The spectrogram obtained from an all-optical measurement of attosecond pulse emission from a system with an ionization potential of 15.8 eV and Lorentzian $\pi$-phase shift in its recombination moment near 52 eV calculated using the SFA. The dashed green and solid red lines depict the optimal relative perturbing field phase for each photon energy for the atom with and without a phase shift in its recombination moment, respectively. (b) The difference between the optimal relative phase for the measurements with and without the transition moment phase shift. A single-cycle driving field with wavelength 1.8 \textmu m and peak intensity of $1 \times 10^{14}$ W/cm$^{2}$ is used and the perturbing field is a second-harmonic with a relative intensity of $10^{-4}$.}
	\label{fig:s08}
\end{figure*}

\clearpage

\begin{figure*}
	\centering
	\includegraphics[width=0.75\textwidth]{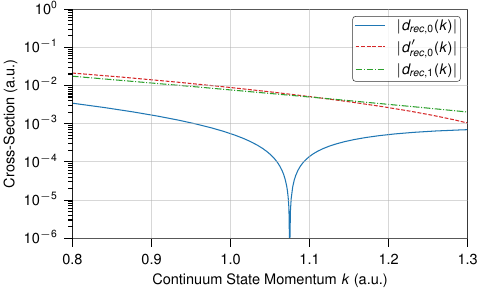}
	\mycaption{\textbf{Recombination Moment Cross-Section in One-Dimensional H$_{2}^{+}$.} }{The magnitudes of the (blue) recombination moment of the ground state $d_{rec, 0}(k)$, (red) its derivative with respect to continuum state momentum $d_{rec,0}'(k)$, and (green) the recombination moment of the first excited state $d_{rec,1}(k)$ of the one-dimensional H$_{2}^{+}$ molecule as a function of the continuum state momentum $k$.}
	\label{fig:s09}
\end{figure*}

\clearpage

\begin{figure*}
	\includegraphics[width=\textwidth]{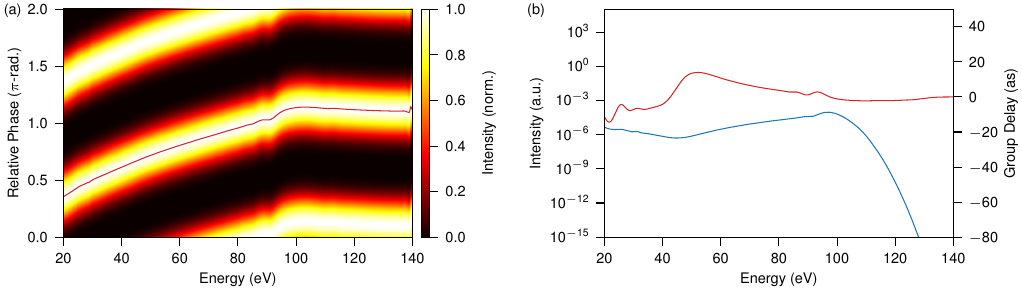}
	\mycaption{\textbf{All-Optical Measurement in a Diatomic Molecule} }{(a) The spectrogram obtained from an all-optical measurement of the one-dimensional diatomic molecule identical to that in \cite{PhysRevA.94.023825}. (b) The (red, right axis) difference between the optimal relative phase for the all-optical measurements of the reference one-dimensional atom and the diatomic molecule and (blue, left axis) the attosecond pulse intensity spectrum from the diatomic molecule. A single-cycle driving field with wavelength 1.8 \textmu m and peak intensity of $1 \times 10^{14}$ W/cm$^{2}$ is used and the perturbing field is a second-harmonic with a relative intensity of $10^{-4}$.}
	\label{fig:s10}
\end{figure*}

\clearpage

\begin{figure*}
	\includegraphics[width=\textwidth]{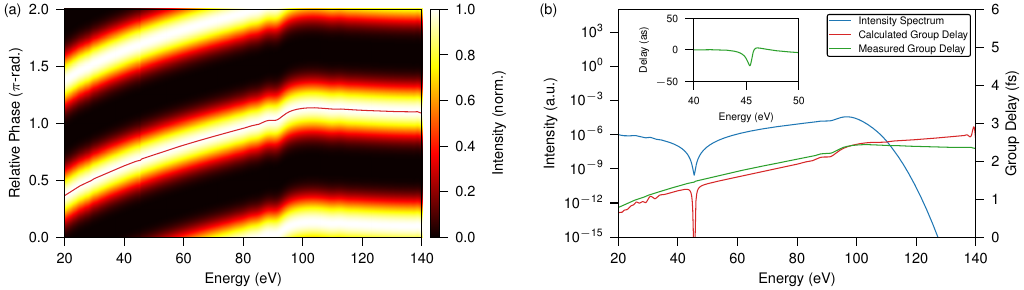}
	\mycaption{\textbf{All-Optical Measurement in a Diatomic Molecule with Projection} }{(a) The spectrogram obtained from an all-optical measurement of the one-dimensional diatomic molecule identical to that in \cite{PhysRevA.94.023825} where we project out the first excited state. The overlaid red line denotes the optimal relative phase between the driving and perturbing fields. (b) The (blue, left axis) attosecond pulse intensity spectrum, (solid red, right axis) calculated group delay, and (solid green, right axis) measured group delay. The inset shows the difference between the measured group delays from a measurement of the diatomic molecule where we project out the first excited state and an equivalent reference atom around the spectral minimum. The fields used for this measurement are identical to those in Fig. \ref{fig:s10}.}
	\label{fig:s11}
\end{figure*}

\clearpage
\nocite{*}
\printbibliography

\end{document}